\documentclass[aps,prl,twocolumn,preprintnumbers,showpacs,nofootinbib]{revtex4}
\usepackage{graphicx}% Include figure files
\usepackage{dcolumn}% Align table columns on decimal point
\usepackage{bm}% bold math
\newcommand{\Od}{{\cal O}}

\newcommand{\beq}{\begin{equation}}
\newcommand{\eeq}{\end{equation}}
\newcommand{\beqa}{\begin{eqnarray}}
\newcommand{\eeqa}{\end{eqnarray}}

\newcommand{\ev}{\text{eV}}

\newcommand{\gev}{\text{GeV}} 

\newcommand{\gr}{\text{G}}
\newcommand{\pla}{\text{Pl}} 
\newcommand{\m}{\text{M}}
\newcommand{\e}{\text{E}}
\newcommand{\eff}{\text{eff}} 

\newcommand{\dm}{\text{DM}}
\newcommand{\loc}{\text{local}}

\newcommand{\sun}{\text{Sun}}
\newcommand{\air}{\text{Air}}

\newcommand{\vp}{\varphi}

\def\beq{\begin{equation}}
\def\eeq{\end{equation}}
\def\bea{\begin{eqnarray}}
\def\eea{\end{eqnarray}}
\newcommand{\be}{\begin{equation}}

\newcommand{\ee}{\end{equation}}
\def\beqa{\begin{eqnarray}}
\def\eeqa{\end{eqnarray}}
\def\beq{\begin{equation}}
\def\eeq{\end{equation}}

\def\gd{g_{\mu\nu}}

\def\dab{_{\alpha\beta}}

\let\gam=w

\def\gd{g_{\mu\nu}}

\renewcommand{\epsilon}{\varepsilon}

\def\ep{\epsilon}
\def\d{\delta}
\def\I{{\cal I}}
\begin{document}

\preprint{UCI-TR-2005-30}  \preprint{gr-qc/0507039}

\input epsf \renewcommand{\topfraction}{0.8}
%\pagestyle{empty}
%\begin{flushright}
%{}
%\end{flushright}
%\vspace*{-25mm}

\title{The Newtonian limit at intermediate energies}

%\author{$\,$}
\author{J. A. R. Cembranos}
%\email{jruizcem@uci.edu}

\affiliation{Department of Physics and Astronomy,
 University of California, Irvine, CA 92697 USA}

\date{\today}% It is always \today, today,

\begin{abstract}
We study the metric solutions for the gravitational equations in Modified Gravity Models (MGMs). In models with negative powers of the scalar curvature, we show that the Newtonian Limit (NL) is well defined as a limit at intermediate energies, in contrast with the usual low energy interpretation. Indeed, we show that the gravitational interaction is modified at low densities or low curvatures. 
\end{abstract}


%04.50.+h Gravity in more than four dimensions, Kaluza-Klein theory, unified field theories; alternative theories of gravity (see also %11.25.Mj Compactification and four-dimensional models) 
%04.25.Nx	Post-Newtonian approximation; perturbation theory; related approximations
%98.80.-k Cosmology (see also section 04 General relativity and gravitation; for origin and evolution of galaxies, see 98.62.Ai; for %elementary particle and nuclear processes, see 95.30.Cq; for dark matter, see 95.35.+d; for superclusters and large-scale structure of %the Universe, see 98.65.Dx) 
%95.35.+d Dark matter (stellar, interstellar, galactic, and cosmological) (see also 95.30.Cq Elementary particle processes; for brown %dwarfs, see 97.20.Vs; for galactic halos, see 98.35.Gi or 98.62.Gq; for models of the early Universe, see 97.10.Fy) 
%98.80.Jk Mathematical and relativistic aspects of cosmology  

\maketitle

%\section{Introduction}

Einstein's General Relativity (GR) describes the gravity field in a very successful way by the metric tensor of the space-time through the Einstein-Hilbert (EH) action. However, this action is non renormalizable and many authors have tried to solve different cosmological and astrophysical puzzles by modifying it, i.e. with MGMs; for example, the introduction of Lagrange densities proportional to $R^n$ with $n>1$ (Large MGMs or LMGMs) leads to Starobinsky inflation \cite{Sta}. In the last years, terms proportional to $R^n$ with $n<1$ (Small MGMs or SMGMs) have been taken into account to explain the present cosmic acceleration without the typical cosmological constant ($n=0$) or dark energy \cite{minusone}. In early times, it is reasonable that terms with $n>1$ can be significant but negligible today; whereas terms with $n<1$ can be neglected in the early Universe but not in the present or late epochs, when the space-time curvature is very small.

The introduction of new scalar curvature dependent terms in the action modifies Einstein's Equations (EEs) in the following way:
\begin{eqnarray}\label{equation1}
&&\left[1+\ep'(R)\right]R_{\mu\nu}-
\frac{1}{2}\left[R+\ep(R)\right]g_{\mu\nu}\nonumber\\
&&%\;\;\;\;\;
+\,\,\I_{\alpha\beta\mu\nu}
\nabla^\alpha\nabla^\beta\left[\ep'(R)\right]
=\frac{T_{\mu\nu}}{M_{\pla}^2}\,,
\end{eqnarray}
where $\I_{\alpha\beta\mu\nu}\equiv
\left(g\dab\gd-g_{\alpha\mu}g_{\beta\nu}\right)$ and
$\ep^{(n)} (R)$ is the n$^{th}$ derivative of $\ep(R)$ with respect to the curvature. We are supposing a small modification of the EH action:
\begin{equation}\label{action1}
S_{\gr} =\frac{1}{2} \int d^4 x \sqrt{-g}M_{\pla}^2\left[R
+\ep(R)\right]\;,
\end{equation}
with the reduced Planck mass $M_{\pla}=(8\pi G_N)^{-1/2}\simeq 2.4 \times 10^{18}~\gev$
(we are adopting the convention: $\hbar=c=1$). We are focusing on SMGMs of the form:
%$M_{\pla}$
\begin{eqnarray}\label{MG}
\ep(R)=\lambda\,\mu^{2-2n}R^{n}
\,,\;\;\;\;\;\lambda=\pm1\,,\;\;\;\;\;n<1\,\,(n\ne 0)\,.
\end{eqnarray}
However, other terms present a similar behavior (as a logarithmic
one) and some results of this work are general for any $\ep(R)$.
To the above gravitational action, we have to add the standard
matter one:
\begin{equation}\label{action2}
S_{\m} =\int d^4 x \sqrt{-g}  {\cal L}_{\m}.
\end{equation}
In the case that there is no matter, due to the symmetries of the vacuum, the solutions are maximally symmetric space-times with a constant curvature scalar $R_0$, which is solution of the following equation:
\begin{eqnarray}\label{solution0}
R_0\left[1-\ep'_0\right]+2\ep_0=0\,,
\end{eqnarray}
where $\ep^{(n)}_0\equiv\ep^{(n)}(R_0)$. It is interesting to note
that without a cosmological constant, $R_0=0$, is not in general a
solution for SMGMs. In fact, a large number of works have
studied such modifications with $\mu\sim H_0\sim 10^{-33}$ eV
to explain the observed cosmic
acceleration \cite{minusone,snlog,extram,ad,palatini}, whose origin could come from
different compactifications of extra dimensions in M-theory
\cite{snMT}.

On the other hand, the third term in the left hand side of
(\ref{equation1}) can produce important instabilities
\cite{Dolgov}, whose solution seems to demand the introduction of
higher order terms ($n>1$) in the scalar curvature \cite{snlog}.
However, in the present literature, it is possible to find three other
problems of SMGMs which are directly related to our present
discussion \cite{chiba, Woodard, Dick}. In this work we show that
these three problems are connected and that their analyses are not
enough to exclude SMGMs. On the contrary, we argue that this
type of models presents a well defined NL at intermediate scales.

The NL in SMGMs was studied in \cite{Dick} through weak field expansions around maximally symmetric vacuum solutions. It means that we can perform a series in the $\ep(R)$ function supposing analyticity in the background solution $R_0$:
$\ep(R)=\ep_0+\ep'_0(\d R)+\ep''_0(\d R)^2/2+...$ with $\d R\equiv R-R_0$. For instance, taking trace in (\ref{equation1}) we obtain a covariant expansion, whose zero order is given by (\ref{solution0}) and the first order by:
\begin{eqnarray}\label{equation1st}
\left[1+\ep'_0-R_0\ep''_0\right]\left(\d R\right)
-3\ep''_0\nabla^\alpha\nabla_\alpha\left(\d R\right)
=-\frac{T}{M_{\pla}^2}\,.
\end{eqnarray}
Here, $T=g^{\mu\nu}T_{\mu\nu}$ is the standard energy momentum tensor associated to (\ref{action2}). On the other hand, if we perform a small perturbation of the background metric $g_{\mu\nu}=g_{0\,\mu\nu}+2\Phi\,\d^0_{\mu}\d^0_{\nu}$,  we can write the linearized fourth order equation for the metric perturbation $\Phi$, and interpret this perturbation as the usual Newtonian potential. In \cite{Dick}, it was shown that we can recover an approximate Poisson equation and the NL if $\ep''_0$ is small enough, i.e. at distances $ r\gg(\ep''_0)^{1/2}$.
But in the SMGMs under study:
\begin{eqnarray}\label{potential2}
\ep''(R)=\lambda n(n-1)(R/\mu^2)^{n-1}/R.
\end{eqnarray}
Models with $n<2$ ($n\neq 0,1$) have typically large values of $\ep''_0$ due to their inverse dependence on the Hubble scale $R_0\sim H_0^2$.

On the other hand, in \cite{Woodard}, the gravitational force which appears in the paradigmatic model ($n=-1$) \cite{minusone} due to a diffuse source in a locally de Sitter background 
was calculated. A linearly growing behavior was found, which is unacceptable because, for instance, it can increase the interaction between the Milky Way and Andromeda by six orders of magnitude \cite{Woodard}. To arrive at this conclusion, the authors also performed a similar expansion and analyzed the first order tensorial equation from (\ref{equation1}) whose trace is given by (\ref{equation1st}).

Finally, in Reference \cite{chiba}, it was shown that these SMGMs are equivalent to Scalar-Tensor Theories (STTs) excluded by Solar System (SS) experiments. The MGM action leads to fourth order equations, which can be studied with the usual EEs if we add a new Scalar Degree of Freedom (SDF)--It is the so called {\it Einstein frame}~\cite{Magnano:bd}--. This new field has the following potential:
\begin{eqnarray}\label{potential}
V_{\vp} &=&\frac
{M_{\pla}^2[R(\vp)\ep'_\vp-\ep_\vp]}
{2[1+\ep'_\vp]^2},
\end{eqnarray}
where $R(\vp)$ is the solution of the equation
$\ep'(R)=\exp(\sqrt{2/3}\vp/M_{\pla})-1$, and
$\ep^{(n)}_\vp=\ep^{(n)}(R(\vp))$. Its mass is related to the second derivative of the potential:
\begin{eqnarray}\label{mass}
m_\vp^2=\frac{d^2V_{\vp}}{d\vp^2}=\frac
{1}
{3\ep''_\vp}-\frac
{R(\vp)[3-\ep'_\vp]+4\ep_\vp}
{3[1+\ep'_\vp]^2}.
\end{eqnarray}
In the vacuum solution, this field satisfies
$R(\vp_0)=R_0\sim \mu^2$, which implies that the typical mass is of
order $m_\vp\sim\mu$. However, if $m_\vp\sim H_0< 10^{-18}$ eV, the
model can be excluded by SS tests, such as the deflection of light
by the Sun, because $\vp$ mediates a new force with a long range
\cite{chiba,Will}.

In conclusion, several authors have detected important problems to the viability of SMGMs. Indeed, these three works are related because they have obtained inappropriate behaviors of the gravitational theory taking into account its vacuum state.

Different solutions have been proposed for some of these three problems \cite{snlog}. For instance, Dick \cite{Dick} has proposed a fine tuning to save the gravitational potential. A model that verifies $\ep''_0=0$ has the correct NL, but even more significantly, this tuning resolves the two other problems: the usual gravitational interaction between galaxies is recovered, at least at the linearized equations; and the SDF has a divergent mass (\ref{mass}). Assuredly, the scalar field is not well defined on the vacuum state. This result can be understood because this field takes into account the new degree of freedom of the gravity due to its fourth order equations, but if $\ep''_0=0$, the metric has associated the standard second order equations in vacuum.

A fine tuning usually means a physical misunderstanding, but there could be a fundamental reason for this cancellation. For instance, Dick has proposed the following example
\cite{Dick}:
\begin{equation}\label{dick}
\ep(R)=-15\mu^4/R+25\mu^6/R^2.
\end{equation}
The vacuum solution reads $R_0=5\mu^2$. These two terms have the
same importance as the EH one in the vacuum state, which means
that the gravitational coupling is modified. The effective
$M_{\pla}$ can be deduced from (\ref{equation1st}):
$M_{\eff}^2=M_{\pla}^2(1+\ep'_0)$, and in this particular model:
$M_{\eff}^2=5M_{\pla}^2/6$ \cite{Dick}.

On the other hand, in Reference \cite{Arvind}, it has been shown that we
can recover Newton's Gravity Law (NGL) through the
Schwarzschild solution inside a de Sitter (or anti-de Sitter)
space. In fact, the metric:
\begin{eqnarray}
ds^2&=&-A(r)dt^2+A(r)^{-1}dr^2+r^2d\Omega^2\,,\nonumber\\
A(r)&=&1-\frac{2m}{M_{\pla}^2r}-\frac{R_0r^2}{12}\,
\end{eqnarray}
is solution of the equations of motion (\ref{equation1}) for a
point-like source of mass $m$ and with $R_0$ given by Equation
(\ref{solution0}). Therefore, we can argue that with a small
enough background curvature any astrophysical test of gravity,
which depends on the Schwarzschild solution and its NL, will be
unaffected by the studied SMGMs.

However, this solution is not completely satisfactory because it seems in contradiction with the previous results. Furthermore, without a well defined NL, we can not identify the $M_{\pla}$ in the action. In the same Reference \cite{Arvind}, it has been commented that in the $\mu\rightarrow 0$ limit, we should recover the standard NL. Assuredly, if we remove the modification in the action, i.e. $\ep(R)\rightarrow 0$, we recover the EH one. However, if we follow the arguments presented in \cite{Dick}, we arrive at the conclusion that the effective $M_{\pla}$ is given by $5M_{\pla}^2/6$ in the model  (\ref{dick}) even in the $\mu\rightarrow 0$ limit. We can find similar surprises in the discussed results of \cite{chiba,Woodard}. Without the tuning, the three problems present more inappropriate behaviors in the EH limit.

In our opinion, the question is that SMGMs modify the gravity at low energies. Therefore, it is consistent that the gravitational interaction was very different at low curvatures or low densities. This is the expected behavior, and the three discussed results agree with it showing that the gravitational interaction is more different for smaller values of $\mu$. The interesting question is in what kind of physical environments we can use their results to contrast with experiments or observations.

As it has been pointed out in \cite{Sotiriou}, the expansions in the
scalar curvature perturbations around the de Sitter vacuum solutions for the analyzed SMGMs (\ref{MG}) present the following form:
\begin{eqnarray}
\ep(R)=\ep_0\left[1+\Od\left(\frac{\d R}{R_0}\right)
+\Od\left(\frac{\d R}{R_0}\right)^2
+...\right]\,.
\end{eqnarray}
The experiments which provide us with the most precise values of NGL and the $M_{\pla}$ are realized on the Earth \cite{torsion} with $\d R/R_0\sim\Od(\rho_{\air}/H_0^2M_{\pla}^2)\sim 10^{27}$ (taking into account the air density
$\rho_{\air}\sim 10^{15}$ eV$^4$), which means that we can not identify the $M_{\pla}$ with the formula given in Reference \cite{Dick}. The expansion can not be truncated inside the SS either; for example, we can roughly estimate the minimal SS curvature by the typical local Dark Matter (DM) value $\rho^{\loc}_{\dm}\sim 10^{-1}$ GeV/cm$^3$ $\sim 10^{-6}$ eV$^4$ \cite{PDB}, which implies $\d R/R_0\sim 10^{6}$.

Indeed, we can use this number as a typical value inside a galaxy. In such a case, we can not estimate the force between galaxies with the result reported in \cite{Woodard}. In fact, as their authors recognized, the linearized approximation
around the vacuum solution inside the source (for example, the Milky Way) is not legitimate. Furthermore, the situation is analogous inside the body which can feel this possible force (such as Andromeda). Therefore, the correct metric solution of these non linear equations, which drives its dynamic, can differ very much from the result found in \cite{Woodard}.

Finally, in the equivalent STT, we can not use the vacuum values
for the scalar field to study its effects in the SS \cite{jok}. Without 
performing any calculations, we can guess that SMGMs could present a
very different gravitational interaction close to the vacuum. The
Reference \cite{chiba} has shown it in a very elegant way, but the
experiments or observations which are used to constrain STTs
are not realized on vacuum. The SS tests restrict severely
STTs if the force mediated by the new scalar field 
has associated a long range. However, 
as in the {\it chameleon} case \cite{jok,chameleon}, 
the characteristics of this
field depend significantly on the local curvature and matter content, to which it is
strongly coupled. The situation is again tricky, and to illustrate it, 
we can estimate the range of this new force inside the Sun supposing a 
Yukawa-type potential exponentially suppressed by $m_\vp$.
If $R\gg\mu^2$, we find by using Equations (\ref{mass}) and (\ref{potential2}):
\begin{eqnarray}\label{mass2}
m_\vp^2=\frac{d^2V_{\vp}}{d\vp^2}\simeq\frac
{1}
{3\ep''_\vp}=\frac
{\lambda\,R}
{3n(n-1)}\left(\frac{R}{\mu^2}\right)^{1-n}\,.
\end{eqnarray}
\interfootnotelinepenalty=10000
For instance, for the case $n=-1$: $m_\vp\sim 10^{12}$ eV ($\d R/R_0\sim 10^{30}$ with $\rho_{\sun}\sim 10^{18}\,\ev^4$), we observe that the range is of order of
$m_\vp^{-1}\sim 10^{-19}$ m and conclude that the scalar field can not produce a force with observable effects out of the Sun (with a radius of $r_S\sim 10^{9}$ m). The situation is similar inside the Earth and all the typical SS gravitational sources \footnote{It is interesting to note that we find a negative square-mass term for the model with $\lambda=n=-1$ \cite{minusone} reproducing the instabilities studied in \cite{Dolgov} from an independent approach.}.

The above estimations show that it is difficult to expect that the truncated
expansions around the vacuum state were able to describe any physical observation.
Practically, this type of expansions can only be linearized for cosmological
studies and only in very recent (or future) times when $\d
R/R_0\sim\Omega_{\m}/\Omega_{\Lambda}\sim 0.3$ \cite{CMBR} (or smaller), but
this is one of the most interesting features of SMGMs. They can
modify the usual gravity at low curvatures and explain the present
cosmology without a dark energy component.

If these models make any sense, they have to recover NGL at
intermediate energies. We can not perform an expansion around the
vacuum state, but we can perform an iterative method to resolve
the Equation (\ref{equation1}). The zero order should be given
by EEs ($\ep(R)\rightarrow 0$):
\begin{eqnarray}\label{EE}
R_{\e\,\mu\nu}-\frac{1}{2}R_{\e}\,g_{\e\,\mu\nu}
=\frac{T_{\mu\nu}}{M_{\pla}^2}\,.
\end{eqnarray}
Therefore, we find the usual relations: $R_{\e}=-T/M_{\pla}^2$ and
$R_{\e\,\mu\nu}=(T_{\mu\nu}-T\,g_{\e\,\mu\nu}/2)/M_{\pla}^2$. To obtain
the first correction to EEs, we can interpret the new terms like
a new source: $\ep^{(n)}_{\e}=\ep^{(n)}(-T/M_{\pla}^2)$. Up to this first
order, we can write:
\begin{eqnarray}\label{equationE0}
R_{1\,\mu\nu}-\frac{1}{2}R_1\,g_{1\,\mu\nu}
=\frac{T_{\mu\nu}+T_{1\,\mu\nu}}{M_{\pla}^2}\,,
\end{eqnarray}
with
\begin{eqnarray}\label{equationE1}
T_{1\,\mu\nu}&=&
\frac{M_{\pla}^2\,\ep_{\e}+T\,\ep'_{\e}}{2}
%\left(\right)
\,g_{\e\,\mu\nu} 
-\ep'_{\e} T_{\mu\nu}
\nonumber\\
&-&M_{\pla}^2\,\I_{\e\,\alpha\beta\mu\nu} \nabla_{\e}^\alpha\nabla_{\e}^\beta\,
\ep'_{\e}\,.
\end{eqnarray}
Here, the metric $g_{\e\,\mu\nu}$ is given by EEs (\ref{EE}) and
$\I_{\e\,\alpha\beta\mu\nu}\equiv \left(g{_{\e}}\dab
g_{\e\,\mu\nu}-g_{\e\,\alpha\mu}g_{\e\,\beta\nu}\right)$. 
This procedure is inappropriate for
conformal matter (T=0) inside the studied SMGMs (\ref{MG}), but it is
very helpful for non relativistic matter without pressure,
$T_{\mu\nu}=\rho\,\d^0_\mu\d^0_\nu$:
\begin{eqnarray}\label{equationE1rho}
T_{1\,\mu\nu}&=&\lambda\,M_{\pla}^2\,\mu^{2}\biggl[
\left(\frac{\rho}{M_{\pla}^2\mu^2}\right)^n 
\left(
\frac{1+n}{2}g_{\e\,\mu\nu}
-n\,\d^0_\mu\d^0_\nu
\right)\nonumber\\
&-&n\mu^{-2}\I_{\e\,\alpha\beta\mu\nu}
\nabla_{\e}^\alpha\nabla_{\e}^\beta
\left(\frac{\rho}{M_{\pla}^2\mu^2}\right)^{n-1} \biggr] \,.
\end{eqnarray}
Clearly, we do not find any modification to EEs for $\ep(R)=0$. Indeed, for a small enough value of
$\mu$, the correction is always negligible ($n<1$). It is interesting to estimate when the maximum value of the new energy contribution ($||T_{1\,\mu\nu}||\equiv$ Max$|T_{1\,\mu\nu}|$)
is much smaller than $\rho$, which implies that NGL is recovered. For instance, if we can neglect the temporal and spatial variations of the source, we find the following relation:
\begin{eqnarray}\label{condition}
||T_{1\,\mu\nu}||\ll \,||T_{\mu\nu}||&\Rightarrow&
\rho\gg \mu^2M_{\pla}^2\,,\;\;\;(n<1)
\,.
\end{eqnarray}
We observe that the same estimations, which were used to show that the vacuum
expansions were incorrectly truncated, allow us to conclude that we have a
well defined NL inside the SS for the proposed models with
$\mu\sim H_0$ ($n<1$). In fact, the two approximations
are opposite in certain sense. This NL works for large densities (or
curvatures) in relation to the typical cosmological constant
scale: $M_{\pla}^2 H_0^2\sim$ $10^{-12}$ eV$^4$ ($H_0^2\sim$ $10^{-66}$
eV$^2$), whereas the linearized vacuum expansions work for small densities
(or curvatures) with respect to the same scale.

It is fair to say that we are not able to assure NGL for the
interesting physical environment studied in \cite{Woodard}. The
Condition (\ref{condition}) can not be satisfied in intergalactic
spaces for $\mu\sim H_0$. Indeed, we think that this is another
very interesting property of SMGMs. They modify NGL at
low densities, which is the expected behavior for $n<1$. For
example, if we neglect the mentioned temporal and spatial variations of
the source, we obtain the following correction for the model with
$n=-1$ \cite{minusone}:
\begin{eqnarray}\label{equationE1rhoexample}
T_{1\,\mu\nu}&=&\lambda
\frac{\mu^{4}M_{\pla}^4}{\rho}\d^0_\mu\d^0_\nu\,.
\end{eqnarray}
From this point of view, we have a new source of gravitation,
which is certainly a DM source since we can only detect an
anomalous metric behavior. It may be interesting to
try to explain galactic dynamic anomalies or even rotation
curves without DM halos, which 
requires the introduction of new unobserved particles
in the standard framework \cite{candy}.
 Similar approaches to general MGMs can be
found in previous works with promising results \cite{extram}.
However, if $||T_{1\,\mu\nu}||>\,||T_{\mu\nu}||$, higher order corrections
have to be taken into account and a numerical calculation seems necessary. 
In this sense, we propose an iterative method by repeating the described procedure up to an
eventual convergence. Alternatively, if we know the metric, we can
infer the necessary source. If this source does not agree with
the observed one, we can deduce a possible function $\ep(R)$ that
can explain the difference. For simplicity, in a first step, we
can study the functions given by Equation (\ref{MG}) parameterized
by $\mu$, $n$ and $\lambda$.

Before concluding this article, we would like to comment 
briefly about SMGMs inside the Palatini formalism, in which the
metric and the connection are taken as independent variables
\cite{palatini}. We
think that our discussion can also clarify several aspects
concerning this approach. 
For instance, it is possible to find  in
the literature different works with different results
about the NL in this formulation  \cite{ar,Sotiriou}. 
In our opinion, it is fundamental to work at intermediate energies,
as it has been performed 
very recently in \cite{Sotiriou}, and not
in the traditional low density approximation.

We have studied the viability of SMGMs, which could be interesting to explain not only the present acceleration of the Universe but also its DM content. The idea is that they deviate from GR at low energies or low curvatures. In contrast, we recover the NL at intermediate scales. More analyses to clarify the possible existence of other kinds of problems have to be performed before postulating them as a serious alternative. Work is in progress in this direction.

{\em Acknowledgments} --- 
We thank A. Rajaraman and F. Takayama for important comments.
This Work is supported in part
by NSF PHY-0239817 grant, BFM 2002-01003 project (DGICYT)
and Fulbright-MEC program.


\begin{thebibliography}{99}

\bibitem {Sta} A.A. Starobinsky, {\it Phys.
Lett. B} {\bf 91} (1980) 99;
A.~Dobado and A.~Lopez, {\it Phys. Rev. D} {\bf 52} (1995)
1895

\bibitem{minusone} S. M. Carroll
 {\it et al.}, 
%V.Duvvuri, M. Trodden and M. S. Turner, 
{\it Phys. Rev. D} {\bf 70} (2004) 043528

\bibitem{snlog}
S. Nojiri and S.D. Odintsov, {\it  Phys. Rev. D} {\bf 68} (2003) 123512,
{\it Gen. Rel. Grav.} {\bf 36} (2004) 1765 and
{\it Mod. Phys. Lett. A} {\bf 19} (2004) 627

\bibitem{extram}
  S. Capozziello {\it et al.},
  %S. Capozziello, S. Carloni and A. Troisi,
  %``Quintessence without scalar fields,''
  astro-ph/0303041;
  %%CITATION = ASTRO-PH 0303041;%%
    %``Reconciling dark energy models with f(R) theories,''
  %and 
  {\it Phys.\ Rev.\ D} {\bf 71} (2005) 043503;
  %S. Carloni, P.K.S. Dunsby, S. Capozziello and A. Troisi,
  %``Cosmological dynamics of R**n gravity,''
  gr-qc/0410046;
  %%CITATION = GR-QC 0410046;%%
  %[arXiv:astro-ph/0501426].
  %%CITATION = ASTRO-PH 0501426;%%
  %S. Capozziello, V.F. Cardone, S. Carloni, A. Troisi
  %``Curvature quintessence matched with observational data,''
  {\it Int.\ J.\ Mod.\ Phys.\ D} {\bf 12} (2003) 1969 and
  %[arXiv:astro-ph/0307018]
  {\it Phys. Lett. A} {\bf 326} (2004) 292
  %``Can higher order curvature theories explain rotation curves of galaxies?,''
  %[arXiv:gr-qc/0404114].
  %and astro-ph/0411114;
  %S. Carloni, P.K.S. Dunsby, S. Capozziello and A. Troisi,
  %``Cosmological dynamics of R**n gravity,''
  %gr-qc/0410046
  %%CITATION = GR-QC 0410046;%%
\bibitem{ad}
%\bibitem{Olmo:2005hc}
  G.J.~Olmo,
  %``Post-Newtonian constraints on f(R) cosmologies in metric formalism,''
  gr-qc/0505135
  %%CITATION = GR-QC 0505135;%%
%\cite{Olmo:2005zr}
%\bibitem{Olmo:2005zr}
  and
  %``The gravity lagrangian according to solar system experiments,''
  gr-qc/0505101;
  %%CITATION = GR-QC 0505101;%%
%\bibitem{Navarro:2005gh}
  I.~Navarro and K.~Van Acoleyen,
  %``On the Newtonian limit of Generalized Modified Gravity Models,''
  gr-qc/0506096;
  %%CITATION = GR-QC 0506096;%%
  J.A.R. Cembranos, in preparation
  
\bibitem{palatini}
%\bibitem{v}
D. Vollick, {\it Class. Quant. Grav.} {\bf 21} (2004) 3813;
%\bibitem{palatini}
E. Flanagan, {\it Class. Quant. Grav.} {\bf 21} (2004) 3817;
G. Allemandi {\it et al.},
%, A. Borowiec and M. Francaviglia,
{\it Phys. Rev. D} {\bf 70} (2004) 043524;
P.$\,$Wang and X.$\,$Meng, {\it Phys. Lett. B}$\,\,${\bf 584} (2004)$\,$1 

\bibitem{snMT} S. Nojiri and S.D. Odintsov,
{\it Phys. Lett. B} {\bf 576} (2003) 5

\bibitem{Dolgov} A.
%D.
 Dolgov and M. Kawasaki, {\it Phys. Lett. B}
{\bf 573} (2003) 1

\bibitem{chiba} T. Chiba, {Phys. Lett. B} {\bf 575} (2003) 1

\bibitem{Woodard} M.E. Soussa and R.P. Woodard,
{\it Gen. Rel. Grav.} {\bf 36} (2004) 855

\bibitem{Dick} R. Dick, {\it Gen. Rel. Grav.}  {\bf 36} (2004) 217

\bibitem{Magnano:bd} G. Magnano and L.M. Sokolowski,
{\it Phys. Rev. D} {\bf 50} (1994) 5039

\bibitem{Will}
C.M. Will, {\it Living Rev. Rel.}  {\bf 4} (2001) 4
%%CITATION = GR-QC 0103036;%%

\bibitem{Arvind}
A.~Rajaraman,
%``Newtonian gravity in theories with inverse powers of R,''
astro-ph/0311160

\bibitem{Sotiriou}
  T.P.~Sotiriou,
  %``The nearly Newtonian regime in Non-Linear Theories of Gravity,''
  gr-qc/0507027
  %%CITATION = GR-QC 0507027;%%

\bibitem{torsion}
J.H. Gundlach and S.M. Merkowitz,
%``Measurement of Newton's Constant Using a Torsion Balance with Angular
%Acceleration Feedback,''
{\it Phys. Rev. Lett.}  {\bf 85} (2000) 2869

\bibitem{PDB}
S.~Eidelman {\it et al.}  {\it Phys. Lett. B} {\bf 592} (2004) 1.
%%CITATION = PHLTA,B592,1;%%

\bibitem{jok} J.A.R. Cembranos {\it et al.}, 
%, A. Rajaraman and F. Takayama,
in progress

\bibitem{chameleon}
  J.~Khoury and A.~Weltman,
  %``Chameleon Fields: Awaiting Surprises for Tests of Gravity in Space,''
  {\it Phys.\ Rev.\ Lett.\ }  {\bf 93} (2004) 171104
  %[arXiv:astro-ph/0309300].
  %%CITATION = ASTRO-PH 0309300;%%
  and
  %``Chameleon cosmology,''
  {\it Phys.\ Rev.\ D} {\bf 69} (2004) 044026
  %[arXiv:astro-ph/0309411].
  %%CITATION = ASTRO-PH 0309411;%%
  

\bibitem{CMBR}
  D. N. Spergel {\it et al.}, 
  %[WMAP Collaboration],
  %``First Year Wilkinson Microwave Anisotropy Probe (WMAP) Observations:
  %Determination of Cosmological Parameters,''
  {\it Astrophys. J. Suppl.} {\bf 148} (2003) 175
%  [arXiv:astro-ph/0302209].

\bibitem{candy}
%
J.R. Ellis {\it et al.},
%, J.S. Hagelin, D.V. Nanopoulos, K.A. Olive and M. Srednicki,
%``Supersymmetric Relics From The Big Bang,''
{\it Nucl.\ Phys.\ B} {\bf 238} (1984) 453;
%
L. Covi {\it et al.},
% J.E. Kim and L. Roszkowski,
%``Axinos as cold dark matter,''
{\it Phys.\ Rev.\ Lett.\ }  {\bf 82} (1999) 4180;
%
G. Servant and T.M.P. Tait,
%``Is the lightest Kaluza-Klein particle a viable dark matter candidate?,''
{\it Nucl.\ Phys.\ B} {\bf 650} (2003) 391;
%
H.C. Cheng {\it et al.},
% J.L. Feng and K.T. Matchev,
%``Kaluza-Klein dark matter,''
{\it Phys.\ Rev.\ Lett.\ } {\bf 89} (2002) 211301;
%[hep-ph/0207125].
%
J.A.R. Cembranos {\it et al.},
%, A. Dobado and A.L. Maroto,
%``Brane-world dark matter,''
{\it Phys.\ Rev.\ Lett.\ } {\bf 90} (2003) 241301,
hep-ph/0507066
%
and 
%``Cosmological and astrophysical limits on brane fluctuations,''
{\it Phys.\ Rev.\ D } {\bf 68} (2003) 103505;
%
%J. Alcaraz {\it et al.}, {\it Phys. Rev. D} {\bf 67}
%(2003) 075010;
%
J.L. Feng {\it et al.},
%, A. Rajaraman and F. Takayama,
%``Superweakly-interacting massive particles,''
{\it Phys.\ Rev.\ Lett.\ } {\bf 91} (2003) 011302
%``SuperWIMP dark matter signals from the early Universe,''
and {\it Phys.\ Rev.\ D } {\bf 68} (2003) 063504 
%and %``Graviton cosmology in universal extra dimensions,''
%Phys.\ Rev.\ D {\bf 68}, 085018 (2003)

\bibitem{ar}
A. Dominguez and D. Barraco, {\it Phys.Rev. D} {\bf 70} (2004) 043505;
%\bibitem{Meng}
X.H. Meng and P. Wang,
%``Gravitational potential in Palatini formulation of modified gravity,''
{\it Gen. Rel. Grav.}  {\bf 36} (2004) 1947;
%%CITATION = GR-QC 0311019;%%
%\bibitem{Olmo:2005hd}
  G.J.~Olmo,
  %``Post-Newtonian constraints on f(R) cosmologies in Palatini formalism,''
  gr-qc/0505136
  %%CITATION = GR-QC 0505136;%%


\end{thebibliography}
\end{document}